\documentclass[pra,twocolumn,shownopacs,amssymb]{revtex4}
\usepackage{graphicx}
\usepackage{amsmath}

\begin{document}

\title{Spin-susceptibility of spin-orbit coupled Fermi superfluids}

\author{M. Iskin}
\affiliation{Department of Physics, Ko\c{c} University, Rumelifeneri Yolu, 
34450 Sar\i yer, Istanbul, Turkey}

\date{\today}

\begin{abstract}

Under the self-consistent mean-field approach for the BCS-BEC crossover 
problem, we derive a closed-form analytical expression for the general spin 
response of noncentrosymmetric Fermi superfluids with arbitrary spin-orbit 
coupling and Zeeman fields. In addition to the paramagnetic, i.e., the Pauli
intra-helicity and Van Vleck type inter-helicity, contributions to the spin-susceptibility 
tensor that have normal-state counterparts, we identify a diamagnetic 
inter-helicity contribution that is unique to the superfluid state. 
Our extensive numerical calculations for the Weyl, Rashba and 
equal Rashba-Dresselhaus spin-orbit couplings illustrate that it is this 
diamagnetic contribution that grows gradually with pairing and cancels precisely 
the Van Vleck contribution away from the BCS regime in general. 

\end{abstract}

\pacs{67.85.Lm, 03.75.Ss, 05.30.Fk, 03.75.Hh}

\maketitle

\section{Introduction}
\label{sec:intro}

Having two-dimensional metals with a weak Rashba spin-orbit coupling 
(SOC) in mind, the BCS theory of the superconducting state without 
inversion symmetry~\cite{edelstein95, gorkov01} was initially developed 
to answer how the lack of spatial reflection symmetry can affect properties 
of a superconductor. One of their main conclusions is that, by lifting the two-fold spin 
degeneracy, and hence, mixing the spin-singlet and spin-triplet components 
in the wave function of the Cooper pairs, SOC gives rise to a nonzero and 
rather high spin response for the ground state of noncentrosymmetric 
superconductors. Since the spin susceptibility is commonly used as a 
means to probe and distinguish the spin-singlet and spin-triplet pairings, 
following the original footsteps~\cite{gorkov01}, the spin response of such 
superconductors and superfluids (SF) has been the main subject of many 
papers with broad interest first in the condensed-matter physics 
community~\cite{yip02, friger04, samokhin07, mineev10, smidman17} 
and then in the cold-atom one~\cite{he12, han12, tang14}.

In this paper, we revisit the spin response of uniform noncentrosymmetric 
Fermi SFs with arbitrary SOC and Zeeman fields, and derive a closed-form
analytical expression for the generalized spin-susceptibility tensor 
$\chi_{ij}$ through the derivative expansion of the mean-field thermodynamic 
potential. Our calculation yields three distinct contributions denoted by
$
\chi_{ij} = \chi_{ij}^{intra} + \chi_{ij(a)}^{inter} + \chi_{ij(b)}^{inter}.
$ 
In addition to the usual paramagnetic Pauli intra-helicity contribution 
$\chi_{ij}^{intra}$ and paramagnetic Van Vleck type inter-helicity contribution
$\chi_{ij(a)}^{inter}$ that have normal-state counterparts, we find a 
diamagnetic inter-helicity contribution $\chi_{ij(b)}^{inter}$ that is unique 
to the SF state. 
By performing extensive numerical calculations for the Weyl, Rashba and 
equal Rashba-Dresselhaus (ERD) SOCs, we illustrate that while the spin 
response is dominated quite strongly by 
$\chi_{ij(a)}^{inter}$ in the BCS limit, it is the diamagnetic contribution 
$\chi_{ij(b)}^{inter}$ that grows gradually with pairing and cancels precisely 
the paramagnetic contribution $\chi_{ij(a)}^{inter}$ away from the BCS regime. 
Thus, the existence of a nonzero ground-state spin response is not truly a 
measure of the SOC induced spin-triplet component of the SF order parameter. 
However, since the $\chi_{ij(b)}^{inter}$ contribution has a strong peak 
in the vicinity of unitarity, observation of its diamagnetic effect is within the 
current reach of recent cold-atom experiments~\cite{wang12, cheuk12, 
williams13, huang16, meng16}, as it accounts for the physical mechanism 
that forms spinless molecules from Cooper pairs in the BEC limit. 

The rest of the paper is organized as follows. Starting with an introductory 
discussion of the non-interacting problem in Sec.~\ref{sec:non}, we first derive 
closed-form analytical expressions for uniform SFs in Sec.~\ref{sec:SF}, 
and then present their numerical analysis in Sec.~\ref{sec:numerics}. The paper 
ends with a brief summary of our conclusions in Sec.~\ref{sec:conc}.

\section{Non-interacting Fermi gases}
\label{sec:non}

In order to gain initial insight into the SF Fermi gases, let us first discuss 
the spin-susceptibility tensor of non-interacting Fermi gases that are described by 
the generic Hamiltonian density
$
H_\mathbf{k} = \xi_{\mathbf{k}} \sigma_0 - \mathbf{Z}_\mathbf{k} \cdot \boldsymbol{\sigma}
$
in momentum ($\mathbf{k}$) and spin ($\sigma$) space. Here,
$
\xi_\mathbf{k} = \epsilon_\mathbf{k} - \mu
$ 
is the usual free-particle dispersion $\epsilon_\mathbf{k} = k^2/(2m)$ 
(in units of $\hbar \to 1$ the Planck constant) shifted by the chemical 
potential $\mu$, and
$
\mathbf{Z}_\mathbf{k} = \mathbf{h} - \mathbf{d}_\mathbf{k}
$
is effectively a $\mathbf{k}$-dependent Zeeman field combining the true 
Zeeman field
$
\mathbf{h} = \sum_i h_i \boldsymbol{\widehat{i}}
$
with the SOC field
$
\mathbf{d}_\mathbf{k} = \sum_i d_\mathbf{k}^i \boldsymbol{\widehat{i}},
$
where $\boldsymbol{\widehat{i}}$ is a unit vector along the $i$ 
direction. In addition, $\sigma_0$ is a $2 \times 2$ identity matrix, and 
$
\boldsymbol{\sigma} = \sum_i \sigma_i \boldsymbol{\widehat{i}}
$
is a vector of Pauli spin matrices in such a way that
$
d_\mathbf{k}^i = \alpha_i k_i
$
corresponds to the Weyl SOC when $\alpha_i = \alpha$ for all $i = \{x,y,z\}$,
a Rashba SOC when $\alpha_z = 0$, and to an ERD SOC when 
$\alpha_y = \alpha_z = 0$. Here, we choose $\alpha \ge 0$ 
without the loss of generality. The resultant energy eigenvalues and eigenstates
can be shortly denoted as
$
H_\mathbf{k} |s \mathbf{k} \rangle = \xi_{s\mathbf{k}}(\mathbf{h}) |s \mathbf{k} \rangle,
$
where 
$
\xi_{s\mathbf{k}}(\mathbf{h}) = \xi_\mathbf{k} + s Z_\mathbf{k}(\mathbf{h})
$ 
with $s = \pm$ and $Z_\mathbf{k}(\mathbf{h}) = |\mathbf{Z}_\mathbf{k}(\mathbf{h})|$ is the helicity spectrum.

Expanding the corresponding thermodynamic potential 
$
\Omega(\mathbf{h}) = - T \sum_{s\mathbf{k}} \ln[1 + e^{-\xi_{s \mathbf{k}}(\mathbf{h})/T}]
$
in powers of $\mathbf{h}$, where $T$ is the temperature (in units of $k_B \to 1$ 
the Boltzmann constant), i.e., 
$
\Omega(\mathbf{h}) = \Omega(\mathbf{0}) - (1/2) \sum_{ij} \chi_{ij} h_i h_j + \cdots,
$
we identify the spin-susceptibility tensor as
$
\chi_{ij}  \overset{\mathbf{h} \to \mathbf{0}}{=} - \partial^2 \Omega/(\partial h_i \partial h_j).
$
For non-interacting Fermi gases, a compact way to express this result is as follows
\begin{align*}
\chi_{ij} &= \frac{1}{2}\sum_{ss'\mathbf{k}} \frac{\mathcal{X}_{s\mathbf{k}}-\mathcal{X}_{s'\mathbf{k}}}
{\xi_{s\mathbf{k}}-\xi_{s'\mathbf{k}}}
\underset{\mathbf{h} \to \mathbf{0}}{\mathrm{Re}}
\left\langle s\left| 
\frac{\partial H_\mathbf{k}}{\partial h_i} \right|s' \right\rangle
\left\langle s' \left| \frac{\partial H_\mathbf{k}}{\partial h_j} \right|s \right\rangle,
\end{align*}
where
$
\mathcal{X}_{s\mathbf{k}} = \tanh[\xi_{s\mathbf{k}}/(2T)]
$
is a thermal factor with $\xi_{s\mathbf{k}} = \xi_\mathbf{k} + sd_\mathbf{k}$
and $d_\mathbf{k} = |\mathbf{d}_\mathbf{k}|$, 
$\mathrm{Re}$ is the real part, and $|s \rangle \equiv |s \mathbf{k} \rangle$ is
adapted for a simpler notation. Here, the derivative notion is implied for 
the $s \to s'$ terms. It order to gain further insight, we separate 
$
\chi_{ij} = \chi_{ij}^{intra} + \chi_{ij}^{inter}
$
into its intraband and interband contributions as follows
\begin{align}
\label{eqn:nonintra}
\chi_{ij}^{intra} &= \frac{1}{4T}\sum_{s\mathbf{k}} 
\frac{d_\mathbf{k}^i d_\mathbf{k}^j}{d_\mathbf{k}^2} \mathcal{Y}_{s\mathbf{k}}, \\
\label{eqn:noninter}
\chi_{ij}^{inter} &= \frac{1}{2}\sum_{s\mathbf{k}}
\frac{\mathcal{X}_{s\mathbf{k}}}{s d_\mathbf{k}}
\left( \delta_{ij} - \frac{d_\mathbf{k}^i d_\mathbf{k}^j}{d_\mathbf{k}^2} \right),
\end{align}
where
$
\mathcal{Y}_{s\mathbf{k}} = \mathrm{sech}^2[\xi_{s\mathbf{k}}/(2T)]
$
is another thermal factor, and $\delta_{ij}$ is a Kronecker delta.
While the intraband term corresponds to the Pauli-paramagnetic contribution, 
the interband term is of the Van Vleck type accounting for the field-induced 
virtual transitions between the helicity bands
~\cite{yip02, friger04, samokhin07, mineev10, smidman17}.
Note that we trivially obtain the usual Pauli expression
$
\chi_{ij} = \delta_{ij} \sum_\mathbf{k} \mathcal{Y}_\mathbf{k} / (2T)
$
of a free Fermi gas when $d_\mathbf{k} = 0$.

Analytical understanding of these intraband and interband contributions in 
the ground state proves to be particularly illuminating for the analysis of 
our numerical results that are presented in Sec.~\ref{sec:numerics} for the 
SF Fermi gases. Thus, by setting $T \to 0$, we obtain
$
\chi_{ij}^{intra} =  \sum_{s\mathbf{k}} d_\mathbf{k}^i d_\mathbf{k}^j 
\delta(\xi_{s \mathbf{k}}) /d_\mathbf{k}^2
$
and
$
\chi_{ij}^{inter} = \sum_{s\mathbf{k}} 
s\theta(\xi_{s\mathbf{k}})/
(\delta_{ij}/d_\mathbf{k} - d_\mathbf{k}^i d_\mathbf{k}^j/d_\mathbf{k}^3),
$
where $\delta(x)$ is the Dirac-delta function and $\theta(x)$ is the 
Heaviside-step function. For instance, these expressions lead to
$
\chi_{ij}^{intra} = \delta_{ij} \theta(\mu+m\alpha^2/2)(N/\epsilon_F) 
[\mu/(2\epsilon_F) + m^2\alpha^2/k_F^2]/\sqrt{\mu/\epsilon_F + m^2\alpha^2/k_F^2} 
$
and
$
\chi_{ij}^{inter} = \delta_{ij} \theta(\mu+m\alpha^2/2) (N/\epsilon_F) 
\sqrt{\mu/\epsilon_F + m^2\alpha^2/k_F^2} 
$
for the Weyl SOC in three dimensions, and
$
\chi_{zz}^{intra} = 0,
$
$
2\chi_{\perp}^{intra} = \theta(\mu) N/\epsilon_F
+ \theta(-\mu)\theta(\mu+m\alpha^2/2) (N/\epsilon_F) / \sqrt{1 + 2\mu/(m\alpha^2)},
$
and
$
\chi_{zz}^{inter} = 2\chi_{\perp}^{inter} = \theta(\mu) N/\epsilon_F 
+ \theta(-\mu)\theta(\mu+m\alpha^2/2) (N/\epsilon_F) \sqrt{1 + 2\mu/(m\alpha^2)} 
$
for the Rashba SOC in two dimensions.
Here, $N = k_F^3 V/(3\pi^2)$ and $k_F^2 A/(2\pi)$ are the total number of particles 
in three and two dimensions, respectively, with $\epsilon_F = k_F^2/(2m)$ the 
corresponding Fermi energy.
Thus, in the $\alpha \to 0^+$ limit, we find
$
\chi_{ij}^{intra} = \delta_{ij} N/(2\epsilon_F)
$
and
$
\chi_{ij}^{inter} = \delta_{ij} N/\epsilon_F 
$
for the Weyl SOC in three dimensions, and
$
\chi_{zz}^{intra} = 0,
$
$
2\chi_{\perp}^{intra} = N/\epsilon_F
$
and
$
\chi_{zz}^{inter} = 2\chi_{\perp}^{inter} = N/\epsilon_F
$
for the Rashba SOC in two dimensions. 
Note that these results are in agreement with the usual expression
$
\chi_{ij} = \delta_{ij} DN/(2\epsilon_F)
$
of a free Fermi gas, i.e., when $\mathbf{d}_\mathbf{k} = 0$, in $D$ dimensions.

Having shown that the interband contribution plays an equally important role in 
recovering the usual Pauli paramagnetism of a free Fermi gas in the 
$\mathbf{d}_\mathbf{k} \to 0$ limit, next we investigate its effects on the SF Fermi gases.

\section{Superfluid Fermi gases}
\label{sec:SF}

Assuming a zero-ranged density-density attraction between $\uparrow$ and 
$\downarrow$ particles in real space, we restrict ourselves to the uniform BCS 
description in $\mathbf{k}$ space that is governed by the mean-field Hamiltonian
$
H_{mf} = (1/2) \sum_\mathbf{k} \Psi_\mathbf{k}^\dagger M_\mathbf{k} \Psi_\mathbf{k}
+ \sum_\mathbf{k} \xi_\mathbf{k} + \Delta^2/U.
$
Here, the spinor operator
$
\Psi_\mathbf{k}^\dagger = (\psi_\mathbf{k}^\dagger  \,\,\, \psi_{-\mathbf{k}})
$
with 
$\psi_\mathbf{k}^\dagger = 
(\psi_{\uparrow \mathbf{k}}^\dagger \,\,\, \psi_{\downarrow \mathbf{k}}^\dagger)
$
creates (annihilates) particles (holes) that are characterized by the following matrix
\begin{align}
\label{eqn:ham}
M_\mathbf{k} = 
\left[
\begin{array}{cc}
  \xi_{\mathbf{k}} \sigma_0 - \mathbf{Z}_\mathbf{k}(\mathbf{h}) \cdot \boldsymbol{\sigma} & \mathrm{i} \Delta \sigma_y \\
  - \mathrm{i} \Delta \sigma_y & - \xi_{\mathbf{k}} \sigma_0 + \mathbf{Z}_{-\mathbf{k}}(\mathbf{h}) \cdot \boldsymbol{\sigma}^* \\
\end{array}
\right],
\end{align}
where the BCS mean-field 
$
\Delta = U \langle \psi_{\uparrow \mathbf{k}} \psi_{\downarrow -\mathbf{k}} \rangle,
$
with $U \ge 0$ the strength of the contact interaction and $\langle \cdots \rangle$ 
the thermal average, is taken as a real parameter without the loss of generality. 
Similar to the non-interacting Fermi gases discussed in Sec.~\ref{sec:non},
one can in principle calculate $\chi_{ij}$ by taking the derivatives of the mean-field 
thermodynamic potential
$
\Omega(\mathbf{h}) = -(T/2) \sum_{\lambda \mathbf{k}} 
\ln[1 + e^{-E_{\lambda\mathbf{k}}(\mathbf{h})/T}]
+ \sum_\mathbf{k} \xi_\mathbf{k} + \Delta^2/U,
$
where $\lambda = \{1,2,3,4\}$ labels the quasiparticle and quasihole energies.
While it is possible to simplify this expression in some special cases, i.e.,
by imposing the particle-hole symmetry 
$
E_{\lambda \mathbf{k}}(\mathbf{h}) = -E_{\lambda',-\mathbf{k}}(\mathbf{h})
$
of the system and the traceless condition 
$
\sum_\lambda E_{\lambda \mathbf{k}}(\mathbf{h}) = 0
$ 
for every $\mathbf{k}$, it is unfortunate that the spectra $E_{\lambda \mathbf{k}}(\mathbf{h})$ 
can not be put into a closed-analytical (yet simple) form in the simultaneous presence 
of arbitrary $\mathbf{h}$ and $\mathbf{d}_\mathbf{k}$ fields. 

To circumvent around this problem, next we make use of the Green's function 
approach which allows us to expand $\Omega(\mathbf{h})$ in powers of 
$\mathbf{h}$ in a systematic way~\cite{he12}. For this purpose, we first note that,
$
\Omega(\mathbf{0}) = -(T/2) \sum_{\ell \mathbf{k}} \ln \det G_{\ell\mathbf{k}}^{-1} 
+ \sum_\mathbf{k} \xi_\mathbf{k} + \Delta^2/U,
$
where
\begin{align*}
G_{\ell \mathbf{k}}^{-1} = 
\left[
\begin{array}{cc}
 (\mathrm{i} \omega_\ell - \xi_{\mathbf{k}}) \sigma_0 - \mathbf{d}_\mathbf{k} \cdot \boldsymbol{\sigma} & \mathrm{i} \Delta \sigma_y \\
  - \mathrm{i} \Delta \sigma_y & (\mathrm{i}\omega_\ell + \xi_{\mathbf{k}}) \sigma_0 - \mathbf{d}_{\mathbf{k}} \cdot \boldsymbol{\sigma}^* \\
\end{array}
\right]
\end{align*}
is the inverse Green's function for the $\mathbf{h} = \mathbf{0}$ problem
~\cite{edelstein95, gorkov01, friger04, samokhin07, mineev10,smidman17, he12}.
Here, $\omega_\ell = 2\pi(\ell + 1/2)T$ is the fermionic Matsubara frequency,
and the quasiparticle and quasihole energies are determined by setting
$
\det G_{\ell\mathbf{k}}^{-1} = [(\mathrm{i}\omega_\ell)^2 - E_{+,\mathbf{k}}^2]
[(\mathrm{i}\omega_\ell)^2 - E_{-,\mathbf{k}}^2]
$
to zero where
$
E_{s \mathbf{k}} = \sqrt{\xi_{s\mathbf{k}}^2 + \Delta^2}
$
with $s = \pm$ are the quasiparticle energies.
Similarly, we can write,
$
\Omega(\mathbf{h}) = -(T/2) \sum_{\ell \mathbf{k}} \ln \det \mathcal{G}_{\ell\mathbf{k}}^{-1} 
+ \sum_\mathbf{k} \xi_\mathbf{k} + \Delta^2/U,
$
where 
$
\mathcal{G}_{\ell\mathbf{k}}^{-1} = G_{\ell\mathbf{k}}^{-1} - \Sigma_\mathbf{h}
$
is the inverse Green's function for the $\mathbf{h} \ne \mathbf{0}$ problem, 
where the particle/hole components of $\Sigma_\mathbf{h}$ are
$
\Sigma_\mathbf{h}^{11} = -\mathbf{h} \cdot \boldsymbol{\sigma},
$
$
\Sigma_\mathbf{h}^{22} = \mathbf{h} \cdot \boldsymbol{\sigma}^*
$
and
$
\Sigma_\mathbf{h}^{12} = \Sigma_\mathbf{h}^{21} = 0.
$
Thus, we can formally expand the thermodynamic potential as
$
\Omega(\mathbf{h}) = \Omega(\mathbf{0}) 
+ (T/2) \sum_{n\ell\mathbf{k}} \mathrm{Tr} (G_{\ell\mathbf{k}} \Sigma_\mathbf{h})^n/n,
$
where $\mathrm{Tr}$ denotes a trace over the particle/hole and spin sectors,
and $n \ge 1$ is a positive integer.
Here, the particle/hole components of the Green's function $G_{\ell\mathbf{k}}$
can be written as follows
\begin{align}
\label{eqn:G11}
G_{\ell\mathbf{k}}^{11} &= \frac{1}{2} \sum_s \frac{\mathrm{i} \omega_\ell + \xi_{s \mathbf{k}}}{(\mathrm{i} \omega_\ell)^2 - E_{s\mathbf{k}}^2} 
\left(\sigma_0 + \frac{\mathbf{d}_\mathbf{k} \cdot \boldsymbol{\sigma}}{s d_\mathbf{k}} \right), \\
\label{eqn:G22}
G_{\ell\mathbf{k}}^{22} &= \frac{1}{2} \sum_s \frac{\mathrm{i} \omega_\ell - \xi_{s \mathbf{k}}}{(\mathrm{i} \omega_\ell)^2 - E_{s\mathbf{k}}^2} 
\left(\sigma_0 - \frac{\mathbf{d}_\mathbf{k} \cdot \boldsymbol{\sigma}^*}{s d_\mathbf{k}} \right), \\
\label{eqn:G12}
G_{\ell\mathbf{k}}^{12} &= \frac{1}{2} \sum_s \frac{-\mathrm{i}\Delta\sigma_y}
{(\mathrm{i} \omega_\ell)^2 - E_{s\mathbf{k}}^2} 
\left(\sigma_0 - \frac{\mathbf{d}_\mathbf{k} \cdot \boldsymbol{\sigma^*}}{s d_\mathbf{k}} \right), \\
\label{eqn:G21}
G_{\ell\mathbf{k}}^{21} &= \frac{1}{2} \sum_s \frac{\mathrm{i}\Delta\sigma_y}
{(\mathrm{i} \omega_\ell)^2 - E_{s\mathbf{k}}^2} 
\left(\sigma_0 + \frac{\mathbf{d}_\mathbf{k} \cdot \boldsymbol{\sigma}}{s d_\mathbf{k}} \right),
\end{align}
and they are obtained through a lengthy but a straightforward algebra, 
coinciding with the available literature~\cite{edelstein95, gorkov01, friger04, samokhin07, mineev10,smidman17, he12}. 
Given the power series expansion of $\Omega(\mathbf{h})$ 
with $\mathbf{h}$, we immediately identify
$
\chi_{ij} \overset{\mathbf{h} \to \mathbf{0}}{=} 
- [T/(2h_i h_j)] \sum_{\ell \mathbf{k}} \mathrm{Tr} (G_{\ell\mathbf{k}} \Sigma_\mathbf{h})^2
$
as the spin-susceptibility tensor for the SF Fermi gases.

Using this prescription, and after some tedious algebra, we eventually obtain 
a rather simple expression for
$
\chi_{ij} = - T \sum_{ss'\ell\mathbf{k}} 
\{[(\mathrm{i} \omega_\ell)^2 + \xi_{s\mathbf{k}}\xi_{s'\mathbf{k}}+\Delta^2]/\det G_{\ell\mathbf{k}}^{-1}\}
\underset{\mathbf{h} \to \mathbf{0}}{\mathrm{Re}}
\langle s| \partial H_\mathbf{k}/\partial h_i |s' \rangle
\langle s'| \partial H_\mathbf{k}/\partial h_j |s \rangle,
$
where $H_\mathbf{k}$ is the non-interacting Hamiltonian density introduced 
in Sec.~\ref{sec:non}. Furthermore, performing the summation over the 
Matsubara frequency $\omega_\ell$, a compact way to express the final 
result is as follows
\begin{align}
\chi_{ij} = \frac{1}{2}\sum_{ss'\mathbf{k}} 
& 
\frac
{
\frac{\mathcal{X}_{s\mathbf{k}}}{E_{s\mathbf{k}}} 
\left( \xi_{s\mathbf{k}} + \frac{2\Delta^2}{\xi_{s\mathbf{k}}+\xi_{s'\mathbf{k}}} \right)
-
\frac{\mathcal{X}_{s'\mathbf{k}}}{E_{s'\mathbf{k}}} 
\left( \xi_{s'\mathbf{k}} + \frac{2\Delta^2}{\xi_{s'\mathbf{k}}+\xi_{s\mathbf{k}}} \right)
}
{\xi_{s\mathbf{k}}-\xi_{s'\mathbf{k}}}
\nonumber \\ 
& \times \underset{\mathbf{h} \to \mathbf{0}}{\mathrm{Re}}
\left\langle s\left| 
\frac{\partial H_\mathbf{k}}{\partial h_i} \right|s' \right\rangle
\left\langle s' \left| \frac{\partial H_\mathbf{k}}{\partial h_j} \right|s \right\rangle,
\label{eqn:SFgen}
\end{align}
where
$
\mathcal{X}_{s\mathbf{k}} = \tanh[E_{s\mathbf{k}}/(2T)]
$
is a thermal factor. Here, the derivative notion is again implied for the $s \to s'$ 
terms following the analogous treatment of the non-interacting case. 
In addition, we again separate 
$
\chi_{ij} = \chi_{ij}^{intra} + \chi_{ij(a)}^{inter} + \chi_{ij(b)}^{inter}
$
into its intraband and interband contributions as follows
\begin{align}
\label{eqn:SFintra}
\chi_{ij}^{intra} &= \frac{1}{4T}\sum_{s\mathbf{k}} 
\frac{d_\mathbf{k}^i d_\mathbf{k}^j}{d_\mathbf{k}^2} \mathcal{Y}_{s\mathbf{k}}, \\
\label{eqn:SFintera}
\chi_{ij(a)}^{inter} &= \frac{1}{2}\sum_{s\mathbf{k}}
\frac{\xi_{s\mathbf{k}} \mathcal{X}_{s\mathbf{k}}}{s d_\mathbf{k} E_{s\mathbf{k}}}
\left( \delta_{ij} - \frac{d_\mathbf{k}^i d_\mathbf{k}^j}{d_\mathbf{k}^2} \right), \\
\label{eqn:SFinterb}
\chi_{ij(b)}^{inter} &= \frac{\Delta^2}{2}\sum_{s\mathbf{k}}
\frac{\mathcal{X}_{s\mathbf{k}}}{s d_\mathbf{k} \xi_\mathbf{k} E_{s\mathbf{k}}}
\left( \delta_{ij} - \frac{d_\mathbf{k}^i d_\mathbf{k}^j}{d_\mathbf{k}^2} \right),
\end{align}
where
$
\mathcal{Y}_{s\mathbf{k}} = \mathrm{sech}^2[E_{s\mathbf{k}}/(2T)]
$
is another thermal factor. Note that we trivially obtain the usual BCS expression
$
\chi_{ij} = \delta_{ij} \sum_\mathbf{k} \mathcal{Y}_\mathbf{k} / (2T)
$
of a SF Fermi gas when $d_\mathbf{k} = 0$. These closed-form analytical 
tensors, that are generalized for an arbitrary SOC field $\mathbf{d}_\mathbf{k}$, 
are the main results of this paper. It can be readily verified that their total 
response tensor $\chi_{ij}$ reproduces all known limits in the cold-atom 
literature, e.g., the isotropic Weyl SOC~\cite{he12} and the $zz$-component 
of the Rashba SOC~\cite{tang14} at $T = 0$.
As discussed in Sec.~\ref{sec:numerics}, while $\chi_{ij(a)}^{inter}$ is 
a paramagnetic contribution, $\chi_{ij(b)}^{inter}$ is a diamagnetic one 
in such a way that their sum eventually cancels each other precisely towards
the molecular gas limit. This competition reveals the physical mechanism that 
forms spinless molecules from Cooper pairs in the BEC limit, which naturally 
exhibit zero spin-susceptibility at any $T$ including their ground state.

More importantly, while Eqs.~(\ref{eqn:SFintra}) and~(\ref{eqn:SFintera}) 
evolve, respectively, from Eqs.~(\ref{eqn:nonintra}) and~(\ref{eqn:noninter}) 
of the non-interacting problem as soon as $\Delta \ne 0$, it is only 
Eq.~(\ref{eqn:SFinterb}) that uniquely contributes to the spin-susceptibility
tensor of SF Fermi gases. 
It is intriguing to note that Eq.~(\ref{eqn:SFinterb}) can be put into precisely 
the same form as the interband contribution to the SF density tensor
$
\rho_{ij}^{inter} = -(m\Delta^2/V) \sum_{s \mathbf{k}} d_\mathbf{k} 
\mathcal{X}_{s\mathbf{k}} g_\mathbf{k}^{ij} / (s \xi_\mathbf{k} E_{s\mathbf{k}})
$
~\cite{iskin17} and the pair mass tensor
$
c_{ij}^{inter} = -(1/4) \sum_{s \mathbf{k}} d_\mathbf{k} 
\mathcal{X}_{s\mathbf{k}} g_\mathbf{k}^{ij} / (s \xi_\mathbf{k} \xi_{s\mathbf{k}})
$
near $T \to T_c$~\cite{iskin18}.
For instance, if the SOC field is of the form $d_\mathbf{k}^i = \alpha_i k_i$
then we can reexpress Eq.~(\ref{eqn:SFinterb}) as
$
\chi_{ij(b)}^{inter} = [\Delta^2/(\alpha_i \alpha_j)] \sum_{s\mathbf{k}}  
d_\mathbf{k} \mathcal{X}_{s\mathbf{k}} g_\mathbf{k}^{ij} / 
(s \xi_\mathbf{k} E_{s\mathbf{k}}),
$
where $g_\mathbf{k}^{ij}$ is the total quantum metric of the helicity bands.
What allows this curious correspondence is that the total quantum metric
$
g_\mathbf{k}^{ij} = [\sum_\ell (\partial d_\mathbf{k}^\ell/\partial k_i) 
(\partial d_\mathbf{k}^\ell/\partial k_j) 
- (\partial d_\mathbf{k}/\partial k_i) (\partial d_\mathbf{k}/\partial k_j)]/(2d_\mathbf{k}^2)
$
of the helicity bands reduces to
$
g_\mathbf{k}^{ij} = \alpha_i \alpha_j (d_\mathbf{k}^2 \delta_{ij} - d_\mathbf{k}^i d_\mathbf{k}^j)
/(2d_\mathbf{k}^4)
$
for the specific case when $d_\mathbf{k}^i = \alpha_i k_i$~\cite{iskin17, iskin18}. 
Thus, we identify 
$
\rho_{ij}^{inter} = -(m/V) \alpha_i \alpha_j \chi_{ij(b)}^{inter}
$
for such SOC fields at any $T$. 
From the quantum geometric point of view, this identification reveals that 
the interband term Eq.~(\ref{eqn:SFinterb}) may be interpreted as the quantum 
metric contribution to the spin-susceptibility tensor, caused by the geometric effects on 
the Cooper pairs. Likewise, the remaining interband term Eq.~(\ref{eqn:SFintera}) 
may also be interpreted as the quantum metric contribution to the spin-susceptibility
tensor, caused by the geometric effects on the single particles.

The existence of a nonzero ground-state spin response at $T = 0$ is not truly 
a direct measure of the SOC induced spin-triplet component of the 
SF order parameter. Our analysis reveals that, unlike the interband term 
Eq.~(\ref{eqn:SFinterb}) for the Cooper pairs, the Van Vleck type interband 
term Eq.~(\ref{eqn:SFintera}) has a nonzero contribution even in the normal 
state no matter if $\Delta = 0$. Indeed, next we show that the paramagnetic 
contribution coming from Eq.~(\ref{eqn:SFintera}) dominates the diamagnetic 
one coming from Eq.~(\ref{eqn:SFinterb}) for most of the parameter regimes 
of experimental interest.

\begin{figure*}[htbp]
\includegraphics[scale=0.6]{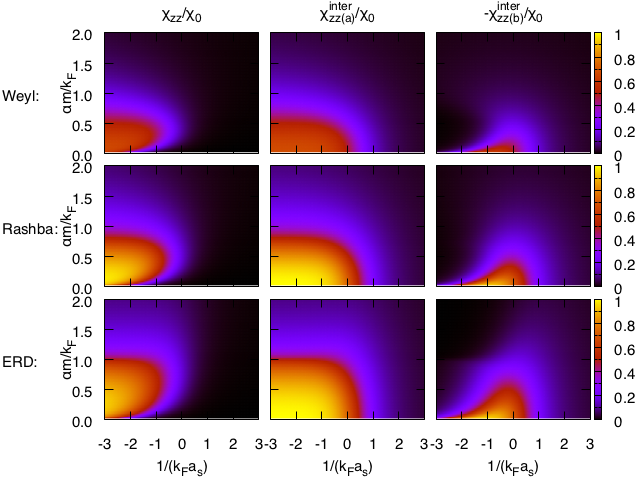}
\caption{(color online)
\label{fig:X3D}
The $zz$-component of the spin-susceptibility tensor is mapped [in units of 
the free Fermi gas value $\chi_0 = 3N/(2\epsilon_F)$] for a three-dimensional 
SF Fermi gas in the ground state. Since $\chi_{zz}^{intra} = 0$ as long as
$\Delta \ne 0$ at $T = 0$, we show $\chi_{zz} = \chi_{zz(a)}^{inter}+\chi_{zz(b)}^{inter}$
together with the competing interband contributions coming from the 
paramagnetic term $\chi_{zz(a)}^{inter}$ and the diamagnetic term 
$\chi_{zz(b)}^{inter}$. Note that this competition gives rise to a cancellation 
of the spin response $\chi_{zz} \to 0$ for the molecular SF Bose gas.
}
\end{figure*}
\section{Numerical Results}
\label{sec:numerics}

It is clear that the numerical evaluations of Eqs.~(\ref{eqn:SFintra}), (\ref{eqn:SFintera})
and (\ref{eqn:SFinterb}) necessitate a priori the solutions of $\Delta$ and $\mu$ for 
any given set of $U$, $\mathbf{d}_\mathbf{k}$ and $T$ parameters. 
Within the BCS mean-field approach, this is achieved through the iterative solutions 
of the self-consistency equations for the SF order parameter
$
1/U = \sum_{s \mathbf{k}} \mathcal{X}_{s \mathbf{k}}/(4VE_{s \mathbf{k}}),
$
and the total number of particles
$
N = \sum_{s \mathbf{k}} [1/2 - \xi_{s\mathbf{k}} \mathcal{X}_{s\mathbf{k}}/(2E_{s\mathbf{k}})].
$
In accordance with the cold-atom literature, while we substitute the theoretical
parameter $U$ with the experimentally more relevant two-body scattering 
length $a_s$ in vacuum via the usual relation
$
V/U = -mV/(4\pi a_s) + \sum_\mathbf{k} 1/(2\epsilon_\mathbf{k})
$
for three-dimensional systems, we substitute $U$ with the two-body 
binding energy $\epsilon_b \ge 0$ in vacuum via the usual relation
$
A/U = \sum_\mathbf{k} 1/(2\epsilon_\mathbf{k} + \epsilon_b)
$
for two-dimensional ones.

Since the mean-field approach works best at low $T$ in the entire BCS-BEC 
crossover regime, next we present our numerical calculations only for the 
SF ground state as it sufficiently illustrates our main findings in this paper. 
For this purpose, we may set $\mathcal{X}_{s\mathbf{k}} \to 1$ 
and $\mathcal{Y}_{s\mathbf{k}} \to 0$ for every $s$ and $\mathbf{k}$ 
as long as $\Delta \ne 0$ in the $T \to 0$ limit, leading to a vanishing 
intraband contribution $\chi_{ij}^{intra} = 0$ for any $\mathbf{d}_\mathbf{k}$. 
Thus, in the SF ground state, while the interband spin-susceptibility tensor 
is isotropic 
$
\chi_{ij}^{inter} = \chi \delta_{ij}
$
for the Weyl SOC~\cite{he12}, its diagonal components are related by
$
\chi_{zz}^{inter} = 2 \chi_\perp^{inter}
$
for the Rashba SOC~\cite{han12}, and by
$
\chi_{xx}^{inter} = 0
$
and
$
\chi_{zz}^{inter} = \chi_{yy}^{inter}
$
for the ERD SOC~\cite{han12} in three dimensions.
However, in two dimensions, the diagonal components are related by
$
\chi_{zz}^{inter} = 2\chi_\perp^{inter}
$
for the Rashba SOC, and by
$
\chi_{xx}^{inter} = 0
$
and
$
\chi_{zz}^{inter} = \chi_{yy}^{inter}
$
for the ERD SOC. Thus, as all of the non-trivial components are proportional 
to each other for all SOC fields considered in this paper, next we present only 
the $zz$-component at $T = 0$.

\begin{figure*}[htbp]
\includegraphics[scale=0.6]{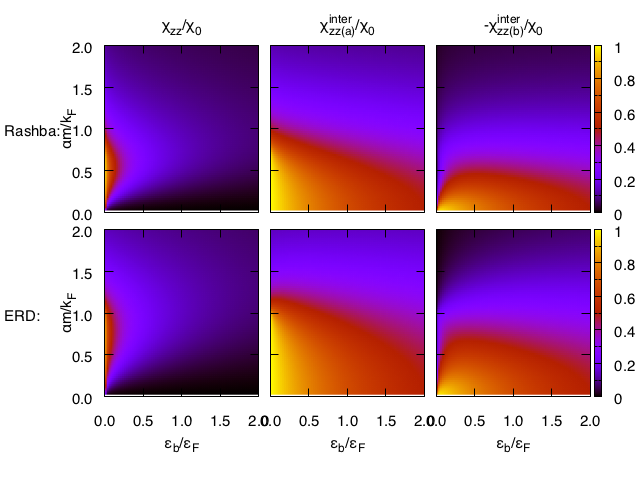}
\caption{(color online)
\label{fig:X2D}
Same as Fig.~\ref{fig:X3D} but for a two-dimensional SF Fermi gas, 
where $\chi_0 = N/\epsilon_F$.
}
\end{figure*}

In Fig.~\ref{fig:X3D}, we show colored maps of
$
\chi_{zz} \equiv \chi_{zz}^{inter} = \chi_{zz(a)}^{inter}+\chi_{zz(b)}^{inter}
$
for the Weyl, Rashba and ERD SOCs introduced in Sec.~\ref{sec:non}, 
along with the specific contributions from $\chi_{zz(a)}^{inter}$ and 
$-\chi_{zz(b)}^{inter}$. See Fig.~\ref{fig:X2D} for the analogous results in two 
dimensions. The strengths $\alpha$ of the SOC fields are varied in both figures 
for the entire BCS-BEC crossover. First, we verify that the peak values of 
$\chi_{zz}$ occuring in the BCS limit are consistent with the $\chi_{zz}^{inter}$ 
contributions that are discussed in Sec.~\ref{sec:non} 
for the non-interacting Fermi gases. 
For instance, $\chi_{zz}^{inter} = N/\epsilon_F$ for the Weyl SOC in three 
dimensions as well as for the Rashba SOC in two dimensions when 
$\alpha \to 0^+$. Second, we observe that $\chi_{zz}$ is dominated by the 
paramagnetic contribution $\chi_{zz(a)}^{inter}$, and never changes sign 
for the entire parameter space. In fact, the magnitude of the diamagnetic 
contribution $|\chi_{zz(b)}^{inter}|$ turns out to be bounded by 
$\chi_{zz(a)}^{inter}$ in the spinless limit of a molecular SF Bose gas in such a 
way that their competition gives rise to the complete cancellation of the 
spin response where $\chi_{zz} \to 0$.

Furthermore, our numerical results clarify the origins of the nonzero spin 
response that is caused by SOC in a variety of noncentrosymmetric 
superconductors and SFs in their ground state. 
Since $\chi_{zz(b)}^{inter}$ is not the leading contribution to $\chi_{zz}$ 
for most of the parameter space, $\chi_{zz} \ne 0$ is clearly not primarily caused 
by the SOC induced spin-triplet component of the SF order parameter. 
In the condensed-matter literature~\cite{gorkov01, yip02, friger04, samokhin07, 
mineev10, smidman17}, Eq.~(\ref{eqn:SFinterb}) is typically neglected and 
Eq.~(\ref{eqn:SFintera}) is replaced with its normal-state value. 
This treatment works reasonably well for weak coupling BCS superconductors 
and SFs, for which Eq.~(\ref{eqn:SFinterb}) has negligible contribution and 
Eq.~(\ref{eqn:SFintera}) approximates the non-interacting result. 
These are clearly illustrated by our extensive numerical results. Away from the 
BCS regime, however, one needs to treat both contributions on an equal footing 
in order to recover the proper molecular limit, which is best seen in Fig.~\ref{fig:X2D}. 

\section{Conclusions}
\label{sec:conc}

In summary, here we analyzed the spin response of spin-orbit coupled Fermi 
SFs with arbitrary SOC and Zeeman fields. For this purpose, assuming a 
self-consistent mean-field approach for the BCS-BEC crossover problem,
we derive a closed-form analytical expression for the generalized 
spin-susceptibility tensor $\chi_{ij}$ through the derivative expansion of 
the thermodynamic potential with Green's function approach. 
The tensor has three distinct contributions denoted as
$
\chi_{ij} = \chi_{ij}^{intra} + \chi_{ij(a)}^{inter} + \chi_{ij(b)}^{inter}.
$ 
In addition to the usual paramagnetic Pauli intra-helicity contribution 
$\chi_{ij}^{intra}$ and paramagnetic Van Vleck type inter-helicity contribution
$\chi_{ij(a)}^{inter}$ that originate from their normal-state counterparts 
upon pairing, we found a diamagnetic inter-helicity contribution 
$\chi_{ij(b)}^{inter}$ that is unique to the SF state. Motivated by the recent 
identification of some interband effects as the quantum metric contributions 
to the SF density and the Cooper pair mass~\cite{iskin17, iskin18}, 
we noted that $\chi_{ij(a)}^{inter}$ and $\chi_{ij(b)}^{inter}$ contributions 
may be interpreted as geometric effects on the single particles and 
Cooper pairs, respectively. However, such a geometric interpretation holds 
only for those SOC fields that are of the generic form 
$
\mathbf{d}_\mathbf{k} = \sum_i \alpha_i k_i \boldsymbol{\widehat{i}}.
$

Furthermore, our extensive numerical calculations for the Weyl, Rashba and 
ERD SOCs illustrated that it is the diamagnetic contribution $\chi_{ij(b)}^{inter}$ 
that grows gradually with pairing and cancels precisely the paramagnetic contribution
$\chi_{ij(a)}^{inter}$ away from the BCS regime. Thus, this competition revealed 
the physical mechanism that forms spinless molecules from Cooper pairs 
in the BEC limit, whose spin response naturally must vanish. 
The existence of a nonzero ground-state spin response is not truly a direct 
measure of the SOC induced spin-triplet component of the SF order parameter. 
Our thorough analysis revealed that, unlike the $\chi_{ij(b)}^{inter}$ 
contribution that is unique to the SF state, 
$\chi_{ij(a)}^{inter}$ contribution is nonzero even in the normal ground state. 
Moreover, we showed that the spin response is dominated quite strongly 
by its normal-state counterpart $\chi_{ij(a)}^{inter}$ for most of the parameter 
regimes of interest. Note that since the $\chi_{ij(b)}^{inter}$ contribution has
a strong peak in the vicinity of unitarity, observation of its diamagnetic effect
is within the reach of ongoing cold-atom experiments~\cite{wang12, cheuk12, 
williams13, huang16, meng16}.

\begin{acknowledgments}
The author acknowledges support from T{\"U}B{\.I}TAK and the BAGEP award 
of the Turkish Science Academy.
\end{acknowledgments}

\end{document}